\documentclass[3p,times,procedia]{elsarticle}
\usepackage{nupha_ecrc}

\volume{00}

\firstpage{1}

\journalname{Nuclear Physics A}

\runauth{David d'Enterria for the CMS Collaboration}

\jid{nupha}

\jnltitlelogo{Nuclear Physics A}

\usepackage{amssymb}
\usepackage[figuresright]{rotating}

\newcommand{\et}{\mathrm{E_{\rm T}}}
\newcommand{\pt}{\mathrm{p_{\rm T}}}

\newcommand{\ptgg}{\mathrm{p}^{\gamma\gamma}_{\rm T}}

\newcommand{\sqrtsnn}{\sqrt{\rm s_{_{\mathrm{NN}}}}}

\newcommand{\epem}{e^+e^-}
\newcommand{\str}{{\sc starlight}} 

\newcommand{\stat}{\,\mathrm{(stat)}}
\newcommand{\syst}{\,\mathrm{(syst)}}
\newcommand{\theo}{\,\mathrm{(theo)}}
\newcommand*{\elm}{e.m.}%\@\xspace}

\begin{document}

\begin{frontmatter}



%% Instructions from Editor: Please use the following \dochead only in the preprint version (e-print arXiv etc.); 
%% use empty \dochead{} when submitting to Nuclear Physics A!
\dochead{XXVIIth International Conference on Ultrarelativistic Nucleus-Nucleus Collisions\\ (Quark Matter 2018)}
%% e.g. \dochead{17th International Conference on Dynamical Processes in Excited States of Solids}

\title{Evidence for light-by-light scattering in ultraperipheral\\ PbPb collisions at $\sqrtsnn$ = 5.02 TeV}


\author{David d'Enterria (for the CMS Collaboration)}

\address{CERN, EP Department, CH-1211 Geneva 23, Switzerland}

\begin{abstract}
Evidence for light-by-light (LbL) scattering, $\gamma\gamma\to\gamma\gamma$, in ultraperipheral PbPb collisions 
at a nucleon-nucleon center-of-mass energy of 5.02 TeV is reported. %in the CMS experiment at the LHC. 
LbL scattering processes are selected in events with just two photons produced, 
with transverse energy E$_{\rm T}^{\gamma}>2$~GeV, pseudorapidity $|\eta^{\gamma}|<2.4$; 
and diphoton invariant mass $m^{\gamma\gamma}>5$~GeV, transverse momentum $\ptgg<1$~GeV, and 
acoplanarity $(1-\Delta \phi^{\gamma\gamma}/\pi)<0.01$. After all selection criteria, 14 events are 
observed, compared to $11.1 \pm 1.1$\,(theo) and $3.8 \pm 1.3$\,(stat) events expected for signal
and background processes respectively. The significance of the signal excess over the background-only hypothesis 
is $4.1\sigma$. The measured fiducial LbL scattering cross section, 
$\sigma_{\mathrm{fid}} (\gamma\gamma\to\gamma\gamma)=122\pm46$\,(stat)\,$\pm29$\,(syst)\,$\pm4$\,(theo)~nb
is consistent with the standard model prediction.
\end{abstract}


\end{frontmatter}

% \linenumbers

%%%%%%%%%%%%%%%%%%%%%%%%%%%%%%%%%%%%%%%%%%%%%%%%%%%%%%%%%%%%%
%% Intro
\section{Introduction}
\label{sec:intro}

\begin{figure*}[hbtp]
\begin{center}
\includegraphics[width=0.99\textwidth,height=4.7cm]{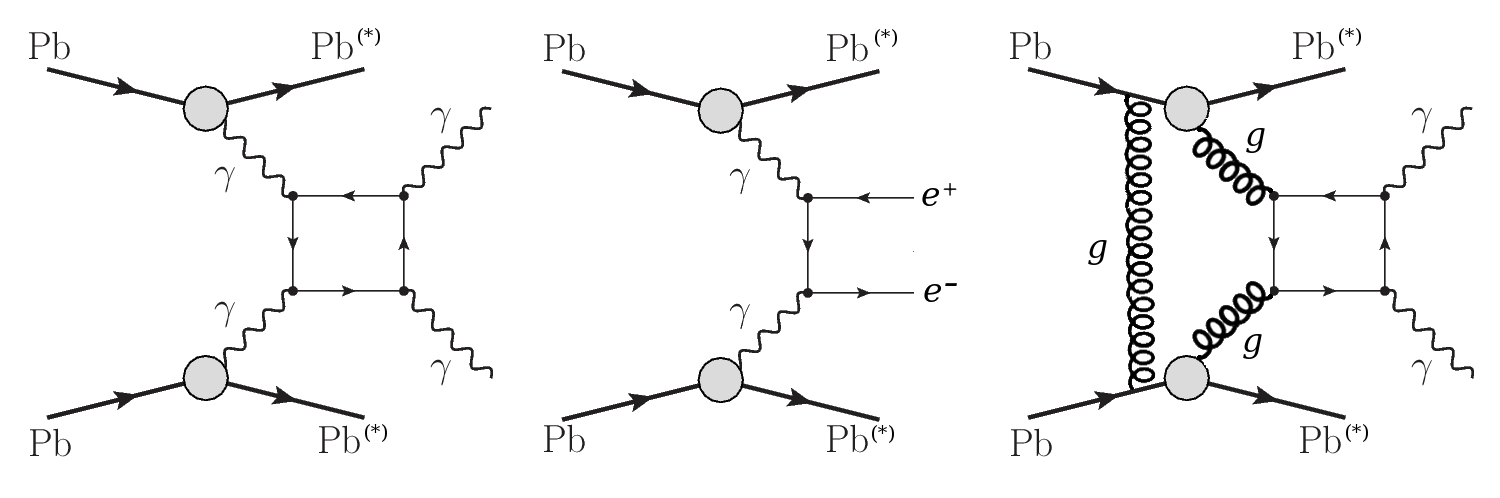}
 \caption{Diagrams of light-by-light scattering ($\gamma \gamma \to \gamma \gamma$, left), QED dielectron 
($\gamma \gamma \to \epem$, center), and central exclusive diphoton ($gg \to \gamma \gamma$, right) production
in ultraperipheral PbPb collisions (with potential electromagnetic excitation$\,^{(*)}$ of the outgoing Pb ions).
\label{fig:feynman}}
\end{center}
\end{figure*}

Elastic light-by-light (LbL) scattering, $\gamma\gamma\to\gamma\gamma$, is a pure quantum mechanical process 
that proceeds, at leading order in the quantum electrodynamics (QED) coupling $\alpha$, via virtual box diagrams 
containing charged particles (Fig.~\ref{fig:feynman}, left). In the standard model (SM), the box diagram involves 
leptons, quarks, and W$^{\pm}$ bosons.
%Although LbL scattering via an electron loop has been indirectly tested through the high-precision measurements 
%of the anomalous electron~\cite{VanDyck:1987ay} and muon~\cite{Brown:2001mga} magnetic moments, its 
Its direct observation in the laboratory remains elusive still today due to a very suppressed production cross 
section proportional to $\alpha^{4}\approx 3\times 10^{-9}$. In Ref.~\cite{d'Enterria:2013yra}, it was proposed
to observe the LbL process at the LHC via ultraperipheral heavy-ion interactions, with impact parameters larger 
than twice the radius of the nuclei, exploiting the very large fluxes of quasi-real photons emitted by the nuclei 
accelerated at TeV energies~\cite{Baltz:2007kq}. 
%The strong electromagnetic fields generated by ultrarelativistic ions can,
%in the equivalent photon approximation (EPA), %~\cite{vonWeizsacker:1934nji,Williams:1934ad,Fermi:1925fq}, 
%be considered as $\gamma$ beams of virtuality $Q^{2} < 1/R^{2}$, where $R$ is the effective radius of the 
%charge distribution. 
For lead (Pb) nuclei with radius $R \approx 7$\,fm, the quasi-real photon beams 
have virtualities $-Q^{2}\approx 1/R < 10^{-3}$\,GeV$^2$, and since each photon flux scales as the square of the ion charge 
$Z^{2}$, $\gamma\gamma$ scattering cross sections in PbPb collisions are enhanced by a factor of $Z^{4}\simeq 5\times 10^{7}$ 
compared to similar proton-proton (pp) or $\epem$ interactions. 
A first evidence of $\gamma\gamma\to\gamma\gamma$ has been reported by the ATLAS experiment~\cite{Aaboud:2017bwk} 
with a signal significance of 4.4$\sigma$ (3.8$\sigma$ expected).
The study of the $\gamma\gamma\to\gamma\gamma$ process at the LHC has also been proposed for searches 
of physics beyond the SM, 
%via \eg{} new heavy particles contributing to the virtual corrections of the box depicted in Fig.~\ref{fig:feynman},
%or new spin-even particles, 
such as axions~\cite{Knapen:2016moh}, %or gravitons~\cite{Sun:2014qba}, appearing as 
%new diphoton resonances on top of it. Also, light-by-light cross sections are sensitive to 
or Born--Infeld extensions of QED~\cite{Ellis:2017edi}.
%, and anomalous quartic gauge couplings~\cite{Chapon:2009hh}. 

The final-state signature of interest is the exclusive production of two photons, 
$\rm PbPb\to\gamma\gamma\to Pb^{(*)}\gamma\gamma Pb^{(*)}$, where the diphoton final state is measured in 
an otherwise empty detector, and the outgoing Pb ions %survive the interaction and 
escape undetected 
%(with a potential electromagnetic excitation denoted by the $^{(*)}$ superscript) 
at very low angles %with respect to the beam 
(Fig.~\ref{fig:feynman}, left). 
The dominant backgrounds are the QED production of an exclusive electron-positron pair ($\gamma\gamma\to\epem$, 
Fig.~\ref{fig:feynman} center) where the $e^\pm$ are misidentified as photons, and gluon-induced 
central exclusive production (CEP)~\cite{Khoze:2004ak} of a pair of photons (Fig.~\ref{fig:feynman}, right).
Simulations of the light-by-light signal are generated with {\sc madgraph v.5}~\cite{Alwall:2011uj} Monte Carlo 
(MC) generator, modified~\cite{d'Enterria:2013yra,dEnterria:2009cwl} to include the nuclear $\gamma$ fluxes and 
the elementary LbL scattering cross section~\cite{Bern:2001dg}. 
%The latter includes all quark and lepton loops at leading order, but omits the 
%{\PW} boson contributions, which are only important for diphoton masses $\mgg > 2 m_{\PW}$.
%Next-to-leading order (NLO) QCD and QED corrections increase $\sigma_{\gamma \gamma \to \gamma \gamma}$
%by just a few percent~\cite{Bern:2001dg} and are neglected here.
%Exclusive $\gamma\gamma\to\epem$ events can be misidentified as LbL scattering if neither electron track 
%is reconstructed or if both electrons undergo hard bremsstrahlung. 
Background QED $\epem$ events are generated with {\sc starlight}~v2.76~\cite{Klein:2016yzr}. %, also based on the EPA fluxes. 
%Since the cross section for the QED $\epem$ background is orders of magnitude larger than that for LbL scattering, and it relies on physics objects
%(electrons) that are very similar to those of the signal (photons), 
%The exclusive dielectron background is analysed in depth in order to estimate many of the (di)photon efficiencies 
%directly from the data, as well as to determine an LbL /(QED $\epem$) production cross sections ratio with reduced 
%common uncertainties.
The CEP process, $gg\to\gamma\gamma$, is simulated with 
{\sc superchic 2.0}~\cite{Harland-Lang:2015cta}, where the computed pp cross section~\cite{Khoze:2004ak} 
is conservatively scaled to the PbPb case by multiplying it by $A^{2}R_{g}^{4}$, where $A$ = 208 is the 
lead mass number and $R_{g} \approx 0.7$ is a gluon shadowing correction in the relevant kinematical 
range~\cite{Eskola:2016oht}, and where the rapidity gap survival factor %, encoding the probability to produce exclusively 
%the diphoton system without any other hadronic activity, 
is taken as 100\%. Given the large theoretical uncertainty of the 
CEP process for PbPb collisions, the absolute normalization of this MC contribution 
is directly determined from a control region in the data. 
%All generated events are passed through the {\sc Geant 4}~\cite{Agostinelli:2002hh} detector simulation. 
%, and the events are reconstructed with the same software as for collision data.
%The simulation describes the tracker material budget with an accuracy better than 10\%, as established by measuring the 
%distribution of reconstructed nuclear interactions and photon conversions in the tracker ~\cite{Chatrchyan:2014fea}.

%%%%%%%%%%%%%%%%%%%%%%%%%%%%%%%%%%%%%%%%%%%%%%%%%%%%%%%%%%%%%
\section{Experimental measurement}
\label{sec:exp}

The measurement is carried out using the following detectors of the CMS experiment~\cite{Chatrchyan:2008zzk}: 
(i) the silicon pixel and strip tracker measures charged particles within pseudorapidities $|\eta|< 2.5$ 
inside the 3.8~T magnetic field, (ii) the lead tungstate crystal electromagnetic calorimeter (ECAL) and a 
brass and scintillator hadron calorimeter (HCAL) reconstruct $\gamma,e^\pm$, and hadrons respectively over 
$|\eta| = 3$, and (iii) the hadron forward calorimeters (HF) measure particle production up to $|\eta| = 5.2$.
Exclusive diphoton candidates are selected with a dedicated level-1 trigger %algorithm
that requires at least two electromagnetic (\elm) clusters with $\et$ above 2~GeV and at least one 
HF detector with total energy below the noise threshold. Offline, 
%events are selected with exactly two photons, each with  $\et > 2$~GeV and $|\eta| <2.4$. 
photons and electrons are reconstructed with the particle flow algorithm~\cite{Sirunyan:2017ulk}.
%global event description (GED)~\cite{Sirunyan:2017ulk}. The GED algorithm 
%allowing for an almost complete recovery of their energy, even if they initiate an \elm{} shower in the material 
%in front of the ECAL. 
In the case of photons, to keep to a minimum the $e^{\pm}$ contamination, 
we require them to be fully unconverted. Additional particle identification (ID) criteria are applied
to remove $\gamma$ from high-$\pt$ $\pi^0$ decays, based on a shower shape analysis. 
%that requires the width of the \elm{} shower along the $\eta$ direction to be below 0.02 (0.06) in the ECAL barrel (endcap). 
Electron candidates are identified by the association of a charged-particle track from the primary vertex
with clusters of energy deposits in the ECAL. Additional $e^\pm$ ID criteria discussed in Ref.~\cite{Chatrchyan:2012tv}
%(isolation, number of tracker hits, HCAL/ECAL energy deposit) 
are applied.
%The electron $E/p$ ratio is found to be within 5\% of unity in the barrel and 15\% in the endcaps. A good agreement 
%is found between data and simulation, both in the $\gamma,e^\pm$ energy scale and resolution.

\begin{figure}[htpb]
\centering
 \includegraphics[width=0.48\textwidth,height=5.7cm]{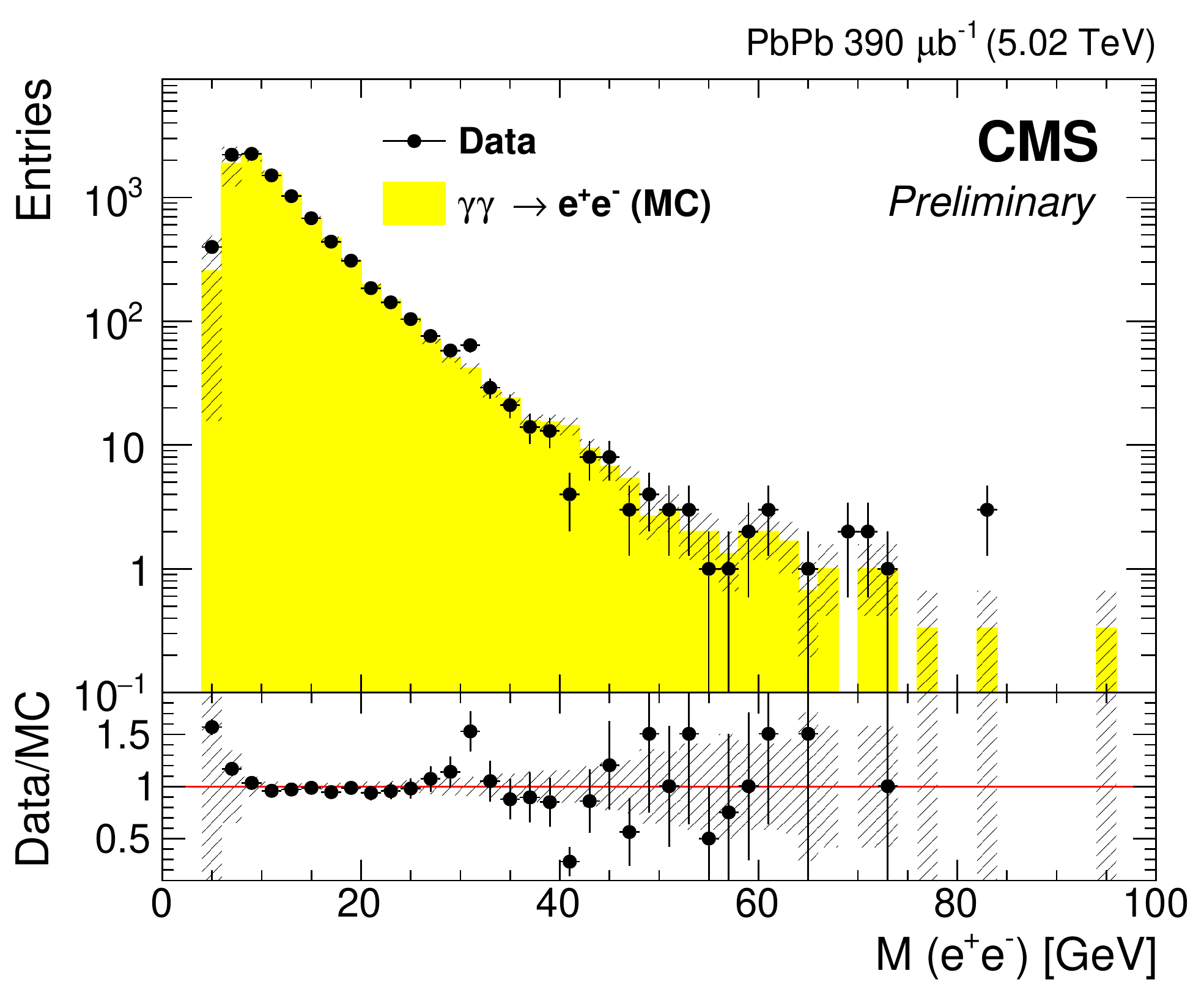}%\hfill
 \includegraphics[width=0.48\textwidth,height=5.7cm]{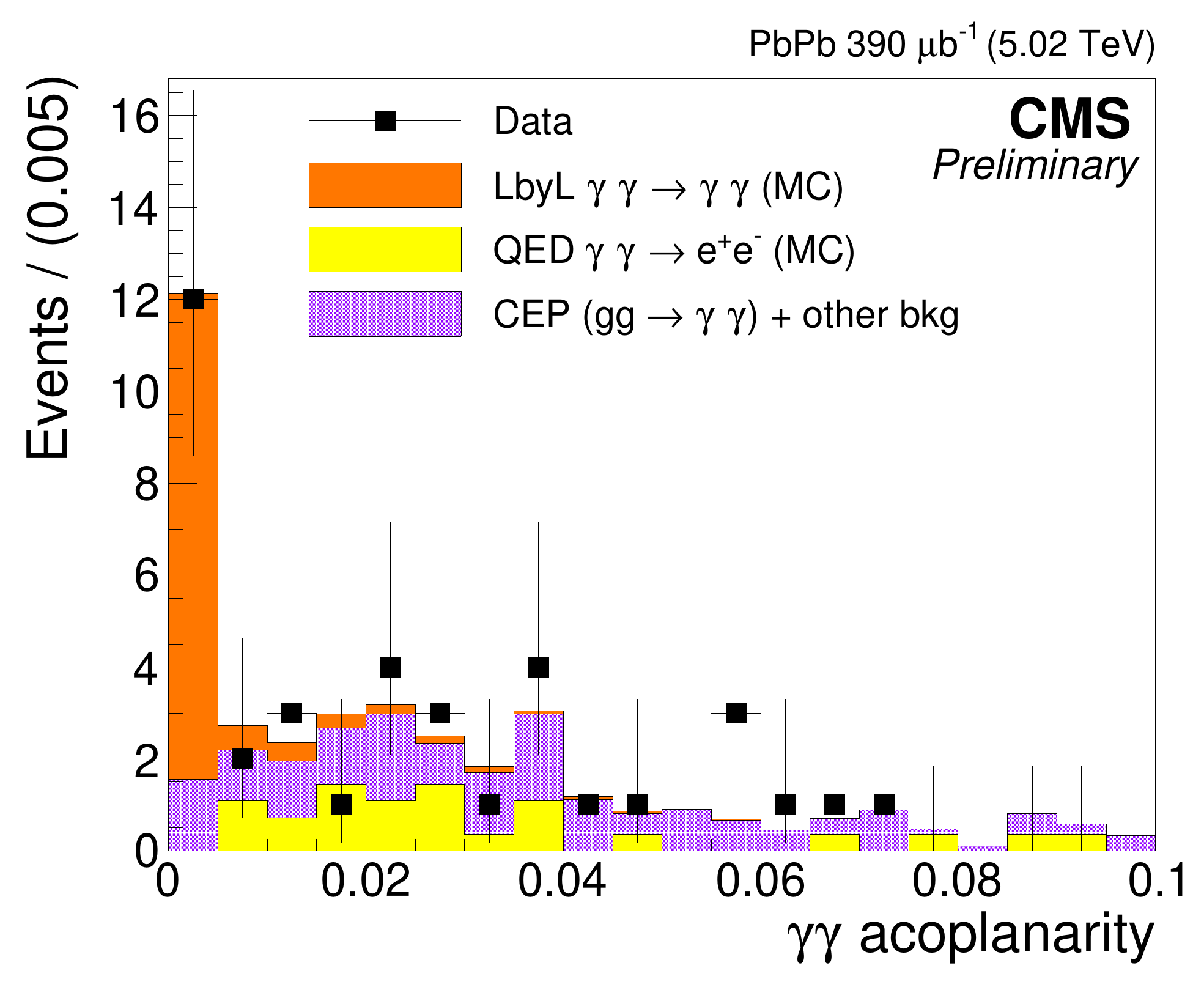}
 \includegraphics[width=0.48\textwidth,height=5.7cm]{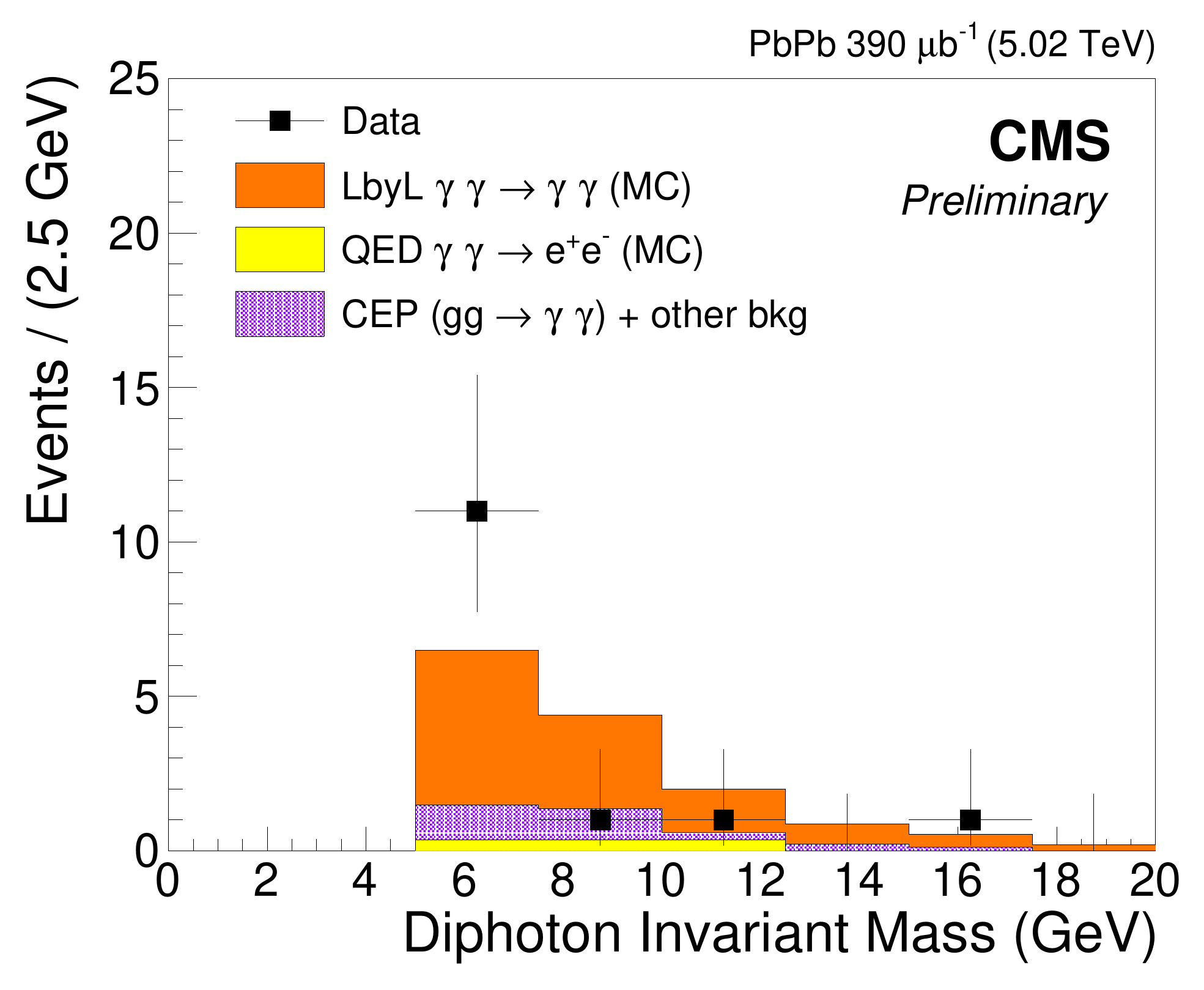}
 \includegraphics[width=0.48\textwidth,height=5.7cm]{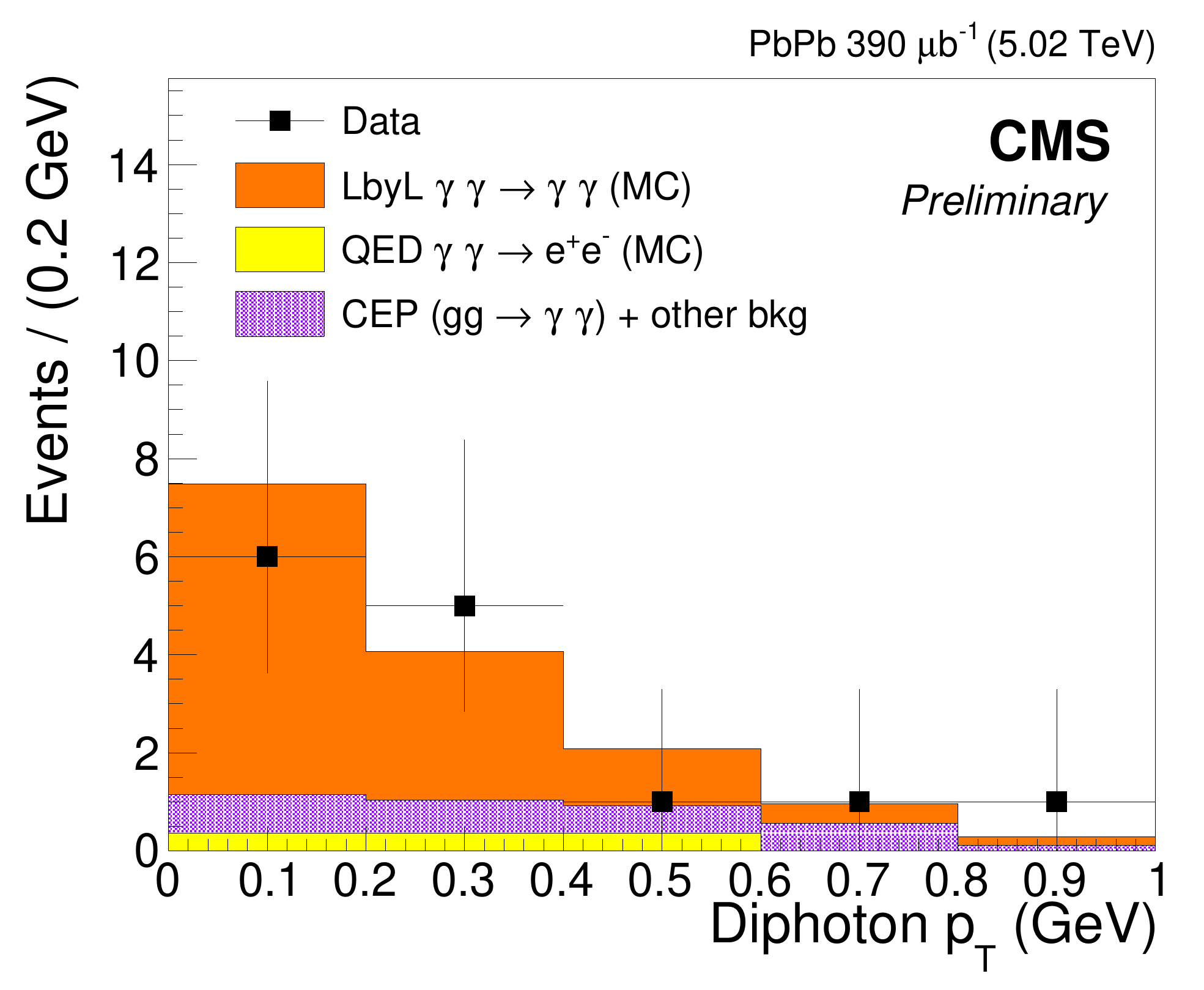}
\caption{\label{fig:QED_LbL} Top-left: Invariant mass distribution for exclusive $\epem$ events in data (circles) 
and \str\ MC expectation (histogram, hashed bands indicate systematic and MC statistical uncertainties  
in quadrature). Top-right: Diphoton acoplanarity for exclusive $\gamma\gamma$ events in data (squares)
compared to the expected LbL scattering signal (orange histogram), QED $\epem$ (yellow histogram), 
and the CEP+other (violet histogram, scaled to match the data in the $\mathrm{A}_{\phi} > 0.02$ region) backgrounds. 
Bottom: Distributions of $m^{\gamma\gamma}$ (left) and $\ptgg$ (right) for exclusive $\gamma\gamma$ events 
in data (squares) and MC (histograms). Error bars indicate statistical uncertainties~\cite{CMS:2018qbh}.
}
\end{figure}

Charged and neutral exclusivity requirements are applied to reject events with any charged particles with 
$\pt > 0.1$~GeV over $|\eta| < 2.4$, and neutral particles above detector noise thresholds over $|\eta| < 5.2$. 
Nonexclusive backgrounds, characterized by a final state with larger 
transverse momenta and larger diphoton acoplanarities, $\rm A_{\phi} = (1-\Delta \phi^{\gamma\gamma}/\pi)$, 
than the back-to-back exclusive $\gamma\gamma$ events, are eliminated by requiring the transverse momentum 
of the diphoton system to be $\ptgg< 1$~GeV, and the acoplanarity of the pair to be 
$\rm A_{\phi} < 0.01$. 
%The two dominant exclusive background sources potentially remaining in the 
%LbL scattering signal region, $\gamma \gamma \to \epem$ and CEP $gg \to \gamma\gamma$, are studied in detail next.
%In order to have a full control of the QED $\epem$ background in the LbL scattering signal region, 
The same analysis carried out for the LbL events is done first on exclusive $\epem$ candidates, 
%applying the same criteria as described above for $\gamma\gamma$ events, 
with the exception that exactly two opposite-sign electrons, instead of exactly two photons, are exclusively reconstructed. 
%, and no additional track with $\pt>0.1$~GeV should be present in addition to the two tracks corresponding to the electrons. 
Figure~\ref{fig:QED_LbL} (top left) shows the acoplanarity distribution measured in exclusive QED $\epem$ events passing 
all selection criteria (circles) compared to the \str\ MC expectation (histogram). 
A good data-simulation agreement is found, thereby confirming the quality of the \elm\
particle reconstruction, of the exclusive event selection criteria, as well as of the UPC MC 
predictions~\cite{d'Enterria:2013yra,Klein:2016yzr}. %for UPC of Pb ions at the LHC. 
Small data--MC differences seen in the $m^{\epem}$ tail
--due to the presence of slightly acoplanar events in data, likely from $\gamma\gamma\to\epem$ events 
where one (or both) electrons radiate an extra soft photon, that are not explicitly simulated--
%. These small discrepancies 
have no impact on the final extracted cross sections integrated over the whole distribution.
The QED dielectron background is then directly estimated from the \str\ MC simulation by counting the number 
of $\epem$ events that pass all LbL scattering selection criteria. The charged exclusivity condition, requiring 
no track in the event above the $\pt = 0.1$~GeV threshold, is successful in removing it almost entirely
(Fig.~\ref{fig:QED_LbL}, top-right).
The simulated CEP $gg \to \gamma\gamma$ events, with large theoretical uncertainties, plus any other residual 
backgrounds resulting in non-fully back-to-back photons %(such as $\gamma\gamma \to \epem \gamma(\gamma)$) 
for which we do not have a simulated sample available, are normalized to match the data in the region $\mathrm{A}_{\phi} > 0.02$, 
where the contribution from $\gamma\gamma\to\gamma\gamma$ is negligible (Fig.~\ref{fig:QED_LbL}, top-right). 
The CEP MC background normalization factor is $ f_\mathrm{nonacoplanar}^\mathrm{norm} = 
% \left(N_\mathrm{data}(\mathrm{A}_{\phi} > 0.02) - N_\mathrm{LbL}^\mathrm{MC}(\mathrm{A}_{\phi} > 0.02) - N_\mathrm{QED}^\mathrm{MC}(\mathrm{A}_{\phi} > 0.02)\right)/\left(N_\mathrm{CEP}^\mathrm{MC}(\mathrm{A}_{\phi} > 0.02)\right) = 
0.95 \pm 0.36\stat$. The final requirement on diphoton acoplanarity ($\mathrm{A}_{\phi} < 0.01$) leads to a significant 
reduction of CEP photon pairs that are produced in diffractive-like processes~\cite{Khoze:2004ak,Harland-Lang:2015cta} 
with larger momentum exchanges, leading to a pair distribution peaking at $\ptgg\approx 0.5$~GeV, and have moderately large tails 
in their azimuthal acoplanarity. %, is achieved with the requirement on diphoton acoplanarity ($\mathrm{A}_{\phi} < 0.01$).
%The number of events due to CEP plus any residual backgrounds is thus estimated to be $3.0 \pm 1.1\stat$. 
%The statistical uncertainties quoted in both values are driven by the size of the data sample 
%left at high acoplanarities, after all selection criteria have been applied.

%Although the LbL and CEP processes share an identical final state, their kinematic distributions are different.
%Diphotons from quasi-real $\gamma\gamma$ fusion processes are produced almost at rest in the transverse plane and, thus, 
%the final-state photons are emitted back-to-back with balanced pair transverse momentum $\pt^{\gamma\gamma}\approx 0$. 

%The exclusive diphoton signal is extracted after applying all selection criteria described above,
%and having estimated the amount of residual QED $\epem$ and CEP+other backgrounds.
After all cuts, we observe 14 LbL scattering candidates, to be compared with $11.1 \pm 1.1\theo$ expected from the LbL scattering signal, 
$2.7\pm 1.1\stat$ from CEP plus any residual $\gamma\gamma$ backgrounds, and $1.1\pm 0.6\stat$ from misidentified QED $\epem$ events. 
Figure~\ref{fig:QED_LbL} (bottom) shows the $\gamma\gamma$ invariant mass (left) and $\ptgg$ (right) measured in data (squares) 
compared to MC expectations. Both the measured total yields and kinematic distributions are in accord with the 
combination of the LbL scattering signal plus exclusive QED $\epem$ and CEP+other backgrounds.
The compatibility of the data with a background-only hypothesis is evaluated from the measured acoplanarity
distribution (Fig.~\ref{fig:QED_LbL}, top-right), using a profile-likelihood ratio as a test statistic.
%including all systematic uncertainties as nuisance parameters with log-normal priors
% ~\cite{Barlow:1993dm,Conway:2011in}. The uncertainty due to the finite size of the MC samples 
% is also included as an additional nuisance parameter for each bin of the histogram. The significance of the excess at low diphoton 
% acoplanarity in data, estimated from the expected distribution of the test statistic for the background-only hypothesis obtained 
% with pseudo-experiments, 
The observed (expected) signal significance is 4.1 (4.4) standard deviations. 
%%%%%%%%%%%%%%%%%%%%%%%%%%%%%%%%%%%%%%%%%%%%%%%%%%%%%%%%%%%%%
%\section{LbL cross section extraction}
%\label{sec:}
%Given the low number of signal counts available for an extraction of differential cross section distributions, 
%An integrated fiducial cross section for LbL scattering above a diphoton mass $m^{\gamma\gamma} = 5$~GeV is calculated. %instead. 
The ratio $R$ of cross sections of the light-by-light scattering over the QED $\epem$ processes is determined
%, thereby reducing the uncertainties related to trigger and reconstruction efficiencies, and integrated luminosity,
%that (partially) cancel out,
from the expression:
\begin{equation}
\label{eq:Req}
 R  =  \frac{\sigma_{\mathrm{fid}} (\gamma\gamma \to \gamma\gamma)}{\sigma(\gamma\gamma \to \epem, m^{\epem}>5\,\mathrm{GeV})} 
 = \frac{N^{\gamma\gamma,\mathrm{data}}-N^{\gamma\gamma,\mathrm{bkg}}}{C^{\gamma\gamma}}
 \times \frac{C^{\epem}\times\mathrm{Acc}^{\epem}}{N^{\epem,\mathrm{data}} \times \mathcal{P}}\,,
\end{equation}
where $\sigma_{\mathrm{fid}} (\gamma\gamma \to \gamma\gamma)$ is the LbL scattering fiducial cross section;
%(passing all the aforementioned $\pt$, $\eta$, $m^{\gamma\gamma}$ kinematic selection criteria for the single $\gamma$ and $\gamma\gamma$ pairs); 
$\sigma(\gamma\gamma \to \epem, m^{\epem}>5$~GeV) is the total QED $\epem$ cross section 
for masses above 5~GeV; $\mathrm{Acc}^{\epem} = N^\mathrm{gen} (\mathrm{p}_{\rm T}^\mathrm{gen} > 2\,\mathrm{GeV}, |\eta^\mathrm{gen}|<2.4, m^{\epem}>5\,\mathrm{GeV}) / N^\mathrm{gen} (m^{\epem}>5\,\mathrm{GeV}) = 0.058 \pm 0.001\stat$  
is the dielectron acceptance for the fiducial single-$e^\pm$ kinematic cuts determined from the \str\ MC generator; 
$N^{\gamma\gamma,\mathrm{data}}$ is the number of diphoton events passing the selection in data; 
$N^{\gamma\gamma,\mathrm{bkg}}$ is the estimated number of background events passing all selection criteria; 
$N^{\epem,\mathrm{data}}$ is the number of dielectron events passing our selection in data; 
$\mathcal{P}$ is the purity of the estimated fraction of QED $\epem$ signal among these dielectron events; and 
$C^{\gamma\gamma}$ and $C^{\epem}$ are overall trigger/reconstruction/exclusivity 
efficiency correction factors, for $\gamma\gamma$ and $\epem$, respectively, 
that are determined from the simulation and confirmed with data-driven studies in control regions.
The ratio of the fiducial LbL scattering to the total QED $\epem$ cross sections 
amounts to $R = (25.4 \pm9.6\stat \pm 6.0\syst) \times 10^{-6}$. 
%where the statistical uncertainty includes in quadrature that from the normalization of the CEP and QED backgrounds in the signal region. 
From the theoretical \str\ prediction of $\sigma(\gamma\gamma \to \epem, m^{ee}>5\mathrm{GeV}) = 4.82 \pm
0.15\theo$~mb, %where the 3\% uncertainty is derived by varying the Pb nucleus
%radius and associated non hadronic-overlap condition in the simulation, 
we finally obtain $\sigma_\mathrm{fid} (\gamma\gamma \to \gamma\gamma) = 122 \pm46\stat \pm29\syst\pm4\theo$~nb,
in good agreement with the theoretical LbL prediction~\cite{d'Enterria:2013yra}  of
$\sigma_\mathrm{fid} (\gamma\gamma \to \gamma\gamma) = 138 \pm 14$~nb in the fiducial region considered.
%The 10\% uncertainty in the LbL theoretical prediction covers different implementations of the non hadronic-overlap 
%condition computed with a Glauber model~\cite{Loizides:2017ack} for varying Pb radius and  
%nucleon-nucleon cross section values, as well as neglected NLO corrections.

%%%%%%%%%%%%%%%%%%%%%%%%%%%%%%%%%%%%%%%%%%%%%%%%%%%%%%%%%%%%%
\section{Summary}
\label{sec:}

Evidence for light-by-light (LbL) scattering, $\gamma \gamma \to \gamma \gamma$, in ultraperipheral PbPb collisions 
at a center-of-mass energy per nucleon pair of 5.02~TeV has been reported. 
%, based on a data sample corresponding to an integrated luminosity of 390\mubinv
%recorded by the CMS experiment at the LHC in 2015. 
Fourteen LbL scattering candidate-events with just two photons produced have been observed passing all kinematical selection 
requirements: single-photon $\mathrm{E}_{\mathrm{T}}^{\gamma}>2~\mathrm{GeV}$, pseudorapidity 
$|\eta^{\gamma}|<2.4$; and diphoton invariant mass $\mathrm{m}^{\gamma\gamma}>5~\mathrm{GeV}$,
transverse momentum $\mathrm{p}_{\mathrm{T}}^{\gamma\gamma}<1~\mathrm{GeV}$, and 
acoplanarity $(1-\Delta \phi^{\gamma\gamma}/\pi)<0.01$. Both the measured total yields and kinematic 
distributions are in accord with the expectations from the LbL scattering signal plus small residual backgrounds, mostly
from misidentified exclusive dielectron, $\gamma \gamma \to\epem$, and gluon-induced central exclusive, 
$gg\to \gamma\gamma$, processes. The observed (expected) significance of the LbL scattering signal over 
the background-only expectation is  4.1 (4.4) standard deviations. The measured fiducial light-by-light 
scattering cross section, $\sigma_\mathrm{fid} (\gamma\gamma \to \gamma\gamma) = 122 \pm46\stat \pm29\syst\pm4\theo$~nb,
is consistent with the standard model prediction. 


\bibliographystyle{elsarticle-num}
%\bibliography{FSQ-16-012}






% \end{thebibliography}

\end{document}